\documentclass[12pt]{article}
\usepackage{amssymb}
\usepackage{amsmath}
\def\dd{{\rm d}}\def\ii{{\rm i}}

\def\half{{\textstyle{\frac12}}}

\def\beq{\begin{equation}}\def\eeq{\end{equation}}
\def\bea{\begin{eqnarray}}\def\eea{\end{eqnarray}}

\begin{document}

\title{Spinor fields in Causal Set Theory}

\author{Roman Sverdlov
\\Physics Department, University of Michigan,
\\450 Church Street, Ann Arbor, MI 48109-1040, USA}
\date{August 21, 2008}
\maketitle

\begin{abstract}

\noindent The goal of this paper is to define fermionic fields on causal set.  This is done by the use of holonomies to define vierbines, and then defining spinor fields by taking advantage of the leftover degrees of freedom of holonomies plus additional scalar fields. Grassmann nature is being enforced by allowing measure to take both positive and negative values, and also by introducing a vector space to have both commutting dot product and anticommutting wedge product. 
\end{abstract}

\noindent{\bf 1. Introduction}

This paper is the fourth paper in a series of papers whose purpose is to define quantum field theory in a causal set. A causal set (for recent review see Ref \cite{causets} ) is a way of discritizing spacetime while retaining strict relativistic invariance. This is done by removing coordinate system, and viewing spacetime as a partially ordered set, where partial ordering is physically viewed as light cone causal relation between different elements of the set, that are viewed as events. Discritization is enforced by the postulate that any interval, $\alpha (p,q) = \{ r \vert p \prec r \prec q \} $ contains only finitely many points.  One of the challenges of the theory is the fact that fields, as they are normally defined, have Lorentzian distances, and so do the derivatives. Hence, in order to be able to do physics in causal set framework, one has to rewrite physics in a way that makes no reference to coordinates. Scalar fields remain to be scalar valued function, vector fields are redefined in terms of holonomies, while gravitational field is redefined in terms of causal relations themselves. In papers  \cite{paper1}, \cite{paper3} and \cite{paper5} a proposed way of estimating Lagrangian densities of these fields were given. The goal of this paper is to do the same for fermionic field. 

This paper is based on "translating" the model of Ref \cite{paper4} from manifold to causal set context. The goal of Ref \cite{paper4} was to re-define fermions in a way that adresses the following two issues:

1)Grassmann nature of the field
2)The presence of vierbines and whether or not they should be viewed as fields

According to this approach, we substitute vierbines with a set of four vector fields that are no longer assumed to be orthogonal to each other.  Extra degrees of freedom that are gained by that are re-interpretted as fermionic field. This implies that, indeed, vierbines are fields since both vierbines as well as spinor fields are different degrees of freedom of the same thing, so if the latter is a field, the former is a field as well. Furthermore, Grassmann nature of fermionic field is enforced by equipping that space with both commutting dot and anticommuting wedge product, together with cleverly defined measure that has both positive and negative values. 

The goal of this paper is to rewrite the results of \cite{paper4} while replacing the four vector fields with four holonomies. The challenge is to be able to express the contractions of vector fields itself as well as their derivatives in a coordinate independent way, based on holonomies alone. 

\noindent{\bf 2. Quick review of \cite{paper4}}

Since this paper is based on results of \cite{paper4} it is best to summarize these results before proceeding with this paper. 

In Ref \cite{paper4} fermions were described in terms of the four orthonormal vector fields, for the case of standard, coordinate-based manifold. It was done based on the following three steps:

PART 1: DEFINE GRASSMANN NUMBERS AS LITERAL MATHEMATICAL OBJECTS

We introduced a space equipped with commutting $\cdot$ product and anticommutting $\wedge$ product, together with measure $\xi$ that has both positive and negative values, in such a way that  $\int (d \theta_1 \wedge . . . \wedge d \theta_n) \cdot (\theta_{a_1} \wedge . . . \wedge \theta_{a_m} ) $ obeys expected properties of Grassmann integral. This allows to define Grassmann numbers in a literal sense.

This can be done by enforcing the following two restrictions on measure $\xi$ : 

\beq \int \xi(x)\, \dd x = 0
\eeq 
\beq 
\int x\, \xi(x)\, \dd x = 1
\eeq

For the purposes of the other two steps, it is convenient to define
\bea \xi (x) = \frac{d \delta (x)}{dx} \eea

which can be altered, for the purposes of rigour, to 

\bea \xi (x) = \frac{a^{1.5}}{\pi^{0.5}} e^{-ax^2} \eea

Expected properties of dot and wedge product are the following: 

\beq
\hat{x} \cdot \hat{x} = 1\eeq
\beq
\hat{y} \cdot (\hat{x} \wedge \hat{y}) = - \hat{y} \cdot (\hat{y} \wedge \hat{x}) = \hat{x}\;.
\eeq

\beq
(\hat{y} \wedge \hat{z}) \cdot (\hat{x} \wedge \hat{y}
\wedge \hat{z}) = \hat{x}
\eeq
\beq
\hat{z} \cdot (\hat{x} \wedge \hat{y} \wedge \hat{z})
= \hat{y} \wedge \hat{z}\;.
\eeq

This, of course, is generalized to the products of larger number of multiples. 

PART 2: COME UP WITH MORE GEOMETRIC DEFINITION OF FERMIONS IN A TOY MODEL WHERE THERE ARE NO GRASSMANN NUMBERS. 

Introduce fields $A^{\mu}$, $B^{\mu}$, $C^{\mu}$ and $D^{\mu}$. Define $f_0^{\mu}$ through $f_3^{\mu}$ to be unit vectors in respective directions of $A^{\mu}$ through $D^{\mu}$:

\bea f_0^{\mu} = \frac{A^{\mu}}{\sqrt{A^{\mu}A_{\mu}}} , \;   f_1^{\mu} = \frac{B^{\mu}} {\sqrt{B^{\mu}B_{\mu}}}, \;  f_2^{\mu} = \frac{C^{\mu}}{\sqrt{C^{\mu}C_{\mu}}}  , \;
f_3^{\mu} = \frac{D^{\mu}}{\sqrt{D^{\mu}D_{\mu}}} \eea

  Use Gramm Schmidt process to define a set of orthonormal vectors at each point,  $e_0^{\mu}$, $e_1^{\mu}$, $e_2^{\mu}$, and $e_3^{\mu}$ that are interpretted as vierbeins:
  
  \bea  e_0^{\mu} (A) = \frac{A^{\mu}}{\sqrt{A^{\nu}A_{\nu}}} \eea

\bea e_1^{\mu} (A, B) = \frac{B^{\mu} - e_0^{\nu}B_{\nu} e_0^{\mu}}{\sqrt{(B^{\rho} - e_0^{\alpha}B_{\alpha} e_0^{\rho})(B_{\rho} - e_0^{\beta}B_{\beta} e_{0 \rho})}} \eea

\bea e_2^{\mu} (A, B, C) = \frac{C^{\mu} - e_0^{\alpha}C_{\alpha} e_0^{\mu}-e_1^{\beta}C_{\beta} e_1^{\mu}}{\sqrt{(C^{\rho} - e_0^{\gamma}C_{\gamma} e_0^{\rho}- e_1^{\delta}C_{\delta} e_1^{\delta})(C_{\rho} - e_0^{\delta}C_{\delta} e_{0 \rho}- e_1^{\epsilon}C_{\epsilon} e_{1 \rho})}} \eea

\bea e_3^{\mu} (A, B, C, D) = \frac{D^{\mu} - e_0^{\alpha}D_{\alpha} e_0^{\mu}-e_1^{\beta}D_{\beta} e_1^{\mu}-e_2^{\gamma}D_{\gamma} e_2^{\mu}}{\sqrt{(D^{\rho} - e_0^{\delta}D_{\delta} e_0^{\rho}- e_1^{\epsilon}D_{\epsilon} e_1^{\rho}-e_2^{\phi}D_{\phi} e_2^{\rho})(D^{\rho} - e_0^{\chi}D_{\chi} e_0^{\rho}- e_1^{\eta}D_{\eta} e_1^{\rho}-e_2^{\xi}D_{\xi} e_2^{\rho})}} \eea

   Introduce two scalar fields $\chi_p$ and $\chi_a$, and then define fermionic field to be a rotation of $\chi_p \, \vert u_1 \rangle + \chi_a \, \vert v_1\rangle$ from the $f_a^{\mu}$-based to $e_a^{\mu}$-based reference frame:

\bea
& & \psi_i(\chi_p , \chi_a , \frac{A}{\vert A \vert}, \frac{B}{\vert B \vert}, \frac{C}{\vert C \vert}, \frac{D}{\vert D \vert}) \\
& & =\big(\exp\big\{ -{\textstyle\frac{\ii}{4}} (\ln (e^{-1} (A,B,C,D) f (A,B,C,D)))_{\mu \nu} \sigma^{\mu\nu} \big\}\, \big)_{ij} \big(\chi_p \delta^j_1 + \chi_a \delta^j_3 \big) \;. 
\nonumber
\eea
   
In other words, while $e_a^{\mu}$ are viewed as vierbines, extra degrees of freedom given by $f_a^{\mu}$  allow us to define fermionic field. At the same time, remembering that we have obtained $e_a^{\mu}$  from $f_a^{\mu}$ by Gramm Schmidt process, the fermionic field is viewed as function of $f_a^{\mu}$ alone: $\psi_i (x) \rightarrow \psi_i ( f_0^{\mu} (x), f_1^{\mu} (x), f_2^{\mu} (x), f_3^{\mu} (x))$

PART 3: COMBINE PARTS 1 AND 2 TO OBTAIN LITERAL DEFINITION OF FERMIONIC FIELDS.

Introduce unit vectors $\hat{r_1}$ through $\hat{r_8}$ remembering that we would like $\hat{r_1}$, $\hat{r_3}$, $\hat{r_5}$ and $\hat{r_7}$ to stand for real parts of $\psi_1$ through $\psi_4$ while $\hat{r_2}$, $\hat{r_4}$, $\hat{r_6}$ and $\hat{r_8}$ to stand for imaginary parts of $\psi_1$ through $\psi_4$ Based on this, define the field 

\bea & & \vec{\psi} (\chi_p, \chi_a, A,B,C,D) = \hat{r_1} Re (\psi_1 (\chi_p, \chi_a, A, B, C,D)) + i \hat{r_2} Im (\psi_1 (\chi_p, \chi_a, A, B, C,D)) \nonumber \\
& & + \hat{r_3} Re(\psi_2 (\chi_p, \chi_a, A,B,C,D))+  i \hat{r_4} Im(\psi_2 (\chi_p, \chi_a, A,B,C,D)) \nonumber \\
& & + \hat{r_5} Re(\psi_3 (\chi_p, \chi_a, A,B,C,D)) + i \hat{r_6} Im (\psi_3 (\chi_p, \chi_a, A,B,C,D)) \nonumber \\
& & + \hat{r_7} Re (\psi_4 (\chi_p, \chi_a, A,B,C,D)) + i \hat{r_8} Im (\psi_4 (\chi_p, \chi_a, A,B,C,D)) \eea

Define a measure $\lambda (f_0^{\mu}, f_1^{\mu}, f_2^{\mu}, f_3^{\mu})$ in such a way that its "derivative" with respect  to transformation $( f_0^{\mu} (x), f_1^{\mu} (x), f_2^{\mu} (x), f_3^{\mu} (x)) \rightarrow \psi$ would give us an expected measure $\xi (\psi_i)$ with both positive and negative values as needed for the definition of Grassmann integral (see part 1) :

\bea & & \lambda (\chi_p , \chi_a , A, B, C, D) =\xi \Big[\psi \Big(\chi_p, \chi_a, \frac{A}{\vert A \vert}, \frac{B}{\vert B \vert}, \frac{C}{\vert C \vert}, \frac{D}{\vert D \vert}\Big)\Big] \times \\ \nonumber
& &  \times lim_{\epsilon \rightarrow 0} \epsilon \mu^{-1} \Big\{ \chi_p ' , \chi_a ', A' , B' , C' , D' \Big\vert \big\vert \psi\Big(\chi_p ', \chi_a ', \frac{A'}{\vert A' \vert}, \frac{B'}{\vert B' \vert},\frac{C'}{\vert C' \vert},\frac{D'}{\vert D' \vert} \Big) \nonumber \\
& & - \psi \Big(\chi_p , \chi_a , \frac{A}{\vert A \vert},\frac{B}{\vert B \vert}, \frac{C}{\vert C \vert} , \frac{D}{\vert D \vert} \Big) \big\vert < \epsilon  \wedge \forall a \Big[\Big(e_a^{\mu} \Big( \frac{A'}{\vert A' \vert},\frac{B'}{\vert B' \vert}, \frac{C'}{\vert C' \vert}, \frac{D'}{\vert D' \vert}\Big) - e_a^{\mu} \Big( \frac{A}{\vert A \vert},\frac{B}{\vert B \vert},\frac{C}{\vert C \vert},\frac{D}{\vert D \vert}\Big) \Big)  \nonumber \\
& & \Big(e_{a\mu} \Big(\frac{A'}{\vert A' \vert}, \frac{B'}{\vert B' \vert},\frac{C'}{\vert C' \vert}, \frac{D'}{\vert D' \vert}\Big)   - e_{a\mu} \Big(\frac{A'}{\vert A' \vert}, \frac{B'}{\vert B' \vert},\frac{C'}{\vert C' \vert}, \frac{D'}{\vert D' \vert} \Big)\Big) \Big] < \epsilon^2 \Big\}  \eea 

Here $\mu$ is a usual Eucledian measure on $\mathbb{R}^{18}$ 

We then use this information to rewrite $\int [d \psi] (. . . ) $ in terms of $\int [df_0^{\mu}][df_1^{\mu}][df_2^{\mu}][df_3^{\mu}] ( . . . )$

The bottom line is that in Ref \cite{paper4} fermionic field theory was re-done in terms of four vector fields $A^{\mu}$, $B^{\mu}$, $C^{\mu}$ and $D^{\mu}$. The goal of this paper is to "translate" this theory into causal set context by replacing these four fields with holonomies, thus removing the Lorentzian index. As with other papers (see \cite{paper1}, \cite{paper3} and \cite{paper5}) the ultimate challenge is to express the Lagrangian, which involves contractions of derivatives, in coordinate-independent way.

\bigskip
\noindent{\bf 2. Computting key terms of the Lagrangian}

\bigskip
\noindent{\bf 2.1 General techniques}

Since in our philosophy there are no such things as coordinates, we will interpret vierbines $e_0$, $e_1$, $e_2$ and $e_3$ as four separate fields. Our Lagrangian is 

\beq
L = \overline{\psi}\, \gamma^\mu\, \partial_\mu \psi + 
\overline{\psi}\, \gamma^a \sigma^{ab}\, \psi\, \big(e^\mu_m e^\nu_a 
(\partial_\mu e_{b\nu} - \partial_\nu e_{b\mu}) + e^\rho_a\,
e^\sigma_b\, \partial_\sigma e^m_\rho \big)\;.
\eeq

All terms of the above Lagrangian can be presented in the following three forms: $A^{\mu}B_{\mu}$ , $A^{\mu} \partial_{\mu} \phi$, and  $A^{\mu} B^{\nu} \partial_{\nu} C_{\mu}$. Thus, our goal is to present each of them as a function of holonomies corresponding to each of these fields:

\bea (A^{\mu}B_{\mu} )(p) \approx f(a,b;p) \eea
\bea A^{\mu} \partial_{\mu} \phi \approx g(a, \phi; p) \eea
\bea(A^{\mu} B^{\nu} \partial_{\nu} C_{\mu})(p) \approx h(a,b,c; p) \eea

 We will use the same technique we have used in Ref \cite{paper5},\cite{paper1},\cite{paper3} for bosons. In these papers, a region called Alexandroff set, $\alpha (p,q) = \{ r \vert p \prec r \prec q \}$, was used in finding the approximation to Lagrangian density. The basic steps were as follows: 

a)For any arbitrary Alexandroff set $\alpha (p,q)$, define ${\cal L} (F, \alpha (p,q))$ to be a proposed approximation of Lagrangian density on the interior of that Alexandroff set.

b)Establish a criteria of "reliability" of each of the Alexandroff sets by finding a way to measure "fluctuations" of the characteristic behavior of a given field in their interio, $Fluc (F, \alpha (p,q))$, and the bigger the fluctuations are the less reliable Alexandroff set is. 

c)For any point $p$ define a point $q(F,p)$ in such a way that it minimizes $Fluc (F, \alpha (p,q))$ under the constraint that $\alpha (p,q)$ has more than certain fixed number of points (the latter is needed in order for statistics to be reliable)

While the above basic steps are the same, there are two very different directions of implimenting them. 

1)Integration approach. This apporach was used in \cite{paper1} and \cite{paper3}. According to this approach, part a is carried out in such a way that ${\cal L} (F, \alpha (p,q))$ will give exact value of Lagrangian density, as long as spacetime is flat and all fields are linear. While this  in itself will be obvious advantage as compared to the next approach, the disadvantage is that, as seen in  \cite{paper1} and \cite{paper3}, in order to enforce such a criteria, we need to define Lagrangian to be a linear combination of two or more terms with non-integer coefficients that are very different from any numbers otherwise recognized. This adjustment is needed in order to make sure that, in locally flat coordinates, we have $a=-b$ in ${\cal L} = aT_{00} + b(T_{11}+ T_{22} + T_{33})$ This immediately raises a question as to how come we have to enforce relativity "by hand" if causal set, being based on $\prec$ alone, should be inherently relativistic. 

2)Maximization/Minimization approach. This is an approach that was used in \cite{paper5} According to this approach, in part a we don't bother making sure that the Lagrangian density we propose actually matches the one expected. Instead, we just take one relevent term and "forget" about any others. What we will find is that after having done part c, the other terms that we dropped will no longer be an issue because we have selected q in such a way that Alexandroff set $\alpha (p,q)$ will pick out a frame in which the omitted terms are $0$.  For example, in case of scalar field, we can focus on its time derivative; in part c we will select a frame in which t axis points in the direction of gradient of scalar field, thus time derivative will be all we need for the Lagrangian density. The advantage of this approach is that we no longer have to adjust coefficients since we only have one term to deal with. The disadvantage is that this makes Lagrangian density dependent on the maximization/minimization procedure as opposed to being universally defined for arbitrary Alexandroff set. 

We will now use both approaches to compute $f$, $g$ and $h$.

 %\newpage
\bigskip
\noindent{\bf 2.2 Maximization/minimization approach for fermions}
 $$
\noindent 
$$

I will introduce holonomies in a way similar to paper \cite{paper3}. Namely, I will associate the 
above vector fields with scalar fields of pairs of points that are 
defined in the following way:
\bea
& &a(r,s) = (s^\mu-r^\mu)\, \big(A_\mu(0) + \half\,(r^\nu + s^\nu)\, 
\partial_\nu A_\mu \big) + O(\tau^3) \nonumber\\
& &b(r,s) = (s^\mu-r^\mu)\, \big(B_\mu(0) + \half\,(r^\nu + s^\nu)\, 
\partial_\nu B_\mu \big) + O(\tau^3) \nonumber\\
& &c(r,s) = (s^\mu-r^\mu)\, \big(C_\mu(0) + \half\,(r^\nu + s^\nu)\, 
\partial_\nu C_\mu \big) + O(\tau^3) \nonumber\\
& &d(r,s) = (s^\mu-r^\mu)\, \big(D_\mu(0) + \half\,(f^\nu + s^\nu)\, 
\partial_\nu D_\mu \big) + O(\tau^3)\;.
\eea
The generic things I would like to compute for causal sets is a term of 
the form $A^{\mu} B^{\nu} \partial_\nu C_\mu$ as well as $A^\mu\,B_\mu$.

Let us first compute gauge latter  terms. If we fix point $p$ and think of $a(p,r)$ and $b(p,r)$ as scalar functions of $r$ alone, then we can use the result of the scalar field section and say
\bea
& &|A| = \bigg\vert \frac{\partial a(p,r)}{\partial r^\mu} \bigg\vert
= \bigg(\frac{k_d}{V_0} \bigg)^{\!1/d}
\min_{V(\alpha(p,q)) \geq V_0} \max_{r,s \in \alpha(p,q)} |a(r,s)|
\\
& &|B| = \bigg\vert \frac{\partial b(p,r)}{\partial r^\mu} \bigg\vert
= \bigg(\frac{k_d}{V_0} \bigg)^{\!1/d}
\min_{V(\alpha(p,q)) \geq V_0} \max_{r,s \in \alpha(p,q)} |b(r,s)|
\\
& &|A+B| = \bigg\vert \frac{\partial (a(p,r)+b(p,r))}{\partial r^\mu} \bigg\vert
= \bigg(\frac{k_d}{V_0} \bigg)^{\!1/d}\!\!\! \min_{V(\alpha(p,q)) \geq V_0}\,
\max_{r,s \in \alpha(p,q)} |a(r,s)+b(r,s)|\;.\qquad
\eea
This tells us that
\bea
& &A^\mu\, B_\mu = \half\, (|A+B|^2 - |A|^2 - |B|^2 \\
& &=\ \half\, \bigg(\frac{k_d}{V} \bigg)^{\!1/d}
\Big(\min_{V(\alpha(p,q))\geq V_0}\max_{r,s \in \alpha(p,q)}|a(r,s)+b(r,s)|\nonumber\\
& &\kern76pt-\ \min_{V(\alpha(p,q)) \geq V_0} \max_{r,s \in \alpha(p,q)} |a(r,s)|
- \min_{V(\alpha(p,q)) \geq V_0} \max_{r,s \in \alpha(p,q)} |b(r,s)|\Big)\;.
\eea
Due to the fact that we seen in scalar section that the same equation applies both to spacelike and timelike gradients of $\phi$, it is clear that the above question would be the same if we were replace $A^\mu$ with one of the spacelike vectors. Thus, 
\bea
& &B^\mu\, C_\mu = \half\, \bigg(\frac{k_d}{V} \bigg)^{1/d} \Big(\min_{V(\alpha(p,q)) \geq V_0}\, \max_{r,s \in \alpha(p,q)} |b(r,s)+c(r,s)| \\
& &\kern106pt-\ \min_{V(\alpha(p,q)) \geq V_0}\, \max_{r,s \in \alpha(p,q)} |b(r,s)| - \min_{V(\alpha(p,q)) \geq V_0}\, \max_{r,s \in \alpha(p,q)} |c(r,s)|\Big)\,.\nonumber
\eea
Now, this can also give us the value of $A^\mu\, \partial_\mu\phi$ Namely, I should just replace in my mind $b(r,s)$ with $\phi(s) - \phi(r)$, which means replacing $B_\mu$ with $\partial_\mu \phi$. Thus, we have
\bea
& &A^\mu\, \partial_\mu\phi = \half\, \bigg(\frac{k_d}{V} \bigg)^{\!1/d} \Big(\min_{V(\alpha(p,q))\geq V_0}\,\max_{r,s\in\alpha(p,q)}|a(r,s)+\phi(s)-\phi(r)|\\
& &\kern108pt-\ \min_{V(\alpha(p,q)) \geq V_0}\, \max_{r,s \in \alpha(p,q)} |a(r,s)| - \min_{V(\alpha(p,q)) \geq V_0}\, \max_{r,s \in \alpha(p,q)} |\phi(s)-\phi(r)|\Big)\,.
\nonumber
\eea
Again, the same equation applies if we replace $A^\mu$ with something spacelike:
\bea
& &B^\mu\, \partial_\mu\phi = \half\, \bigg(\frac{k_d}{V} \bigg)^{1/d}
\Big(\min_{V(\alpha(p,q))\geq V_0}\,\max_{r,s\in\alpha(p,q)}|b(r,s)+\phi(s)-\phi(r)|\\
& &\kern108pt-\ \min_{V(\alpha(p,q)) \geq V_0}\,\max_{r,s \in \alpha(p,q)} |b(r,s)| - \min_{V(\alpha(p,q)) \geq V_0}\, \max_{r,s \in \alpha(p,q)} |\phi(s)-\phi(r)|\Big)\,.
\nonumber
\eea
Now let us try to get $A^\mu\, B^\nu\, \partial_\nu C_\mu$.

Suppose we would like to maximize the fluctuation of $a(r,s)b(r,s)c(r,s)d(r,s)$ Ideally, if we could minimize the fluctuation of each of individual four multiples, it would automatically minimize the fluctuation of the product. I claim that we can do just that! Based on the results of scalar part, if gradient of $\phi$ is timelike, in order for fluctuations of $\phi$ to be minimized it has to be parallel to the axis of Alexandroff set; on the other hand, if gradient is spacelike then in order for fluctuations of $\phi$ to be minimized, gradient should lie on equator of Alexandroff set. This means that in order for fluctuations of $a(r,s)$ to be minimized, $A^\mu$ should point along the axis of Alexandroff set, while in order for fluctuations of $b(r,s)$, $c(r,s)$ and $d(r,s)$ to be minimized, then $B^\mu$, $C^\mu$ and $D^\mu$ should lie on equator of Alexandroff set. Now, the amaising thing is that orthogonality condition tells us that, not only these four statements are compatible, but in fact if the condition about the timelike vector $A^\mu$ is met, it {\em forces\/} the conditions about the three spacelike vectors $B^\mu$, $C^\mu$ and $D^\mu$ to be met as well! This means that if we minimize the fluctuations of a product $a(r,s)\,b(r,s)\,c(r,s)\,d(r,s)$ we would also minimize the fluctuations of each of the four multiples individually, which means that we would completely specify the paramenters of my Alexandroff set.

Now suppose we would like to select points $r_1$, $r_2$, $r_3$ and $r_4$ in such a way that minimizes fluctuations of $a(r_1,r_3)\, b(r_2,r_4)$. Again, we can separately maximize each of these two multiplets. To maximize $a(r_1, r_3)$ we have to set $r_1 = p = (-\half\,\tau,0,0,0)$ and $r_3 = q = (\half\,\tau,0,0,0)$. In order to maximize $b(r_2,r_4)$ we have to select $r_2 = (0,\half\,\tau,0,0)$ and $r_4 = (0,-\half\,\tau,0,0,0)$ These two conditions are comparable with each other. Now suppose that we instead decided to maximize fluctuations of $b(r_1,r_3)\, c(r_2,r_4)$ Again, we can simultaneously maximize fluctuations of each of these fields. This time, since both of these fields have spacelike gradient, each one will be maximized by selecting points on the equator. In particular, to maximize $b(r_1,r_3)$ we set $r_1 = (0,-\half\,\tau(p,q),0,0)$ and $r_3 = (0,\half\,\tau(p,q),0,0)$ and in order to maximize $c(r_2, r_4)$ we set $r_2 = (0,0,\half\,\tau(p,q),0)$ and $r_4 = (0,0,-\half\,\tau(p,q),0)$ Again, we get the same square look as we had in gauge case. We notice that the square loop for the case of one timelike and one spacelike field is the same as the square loop for two spacelike fields, except that $t$ axis was replaced by the $y$ axis. So, we will save ourself time and say that in order to maximize $u(r_1,r_3)\, v(r_2,r_4)$ we have to set $U^\mu\, r_{1\mu} = -\half\,\tau$, $U^\mu\, r_{3\mu} = \half\,\tau$, $V^\mu\, r_{2\mu} = \half\,\tau$, $V^\mu\, r_{4\mu} = -\half\,\tau$ regardless of whether the fields are spacelike or timelike. 

Now if $w$ is a third holonomy, then based on our results from gauge part, remembering that all vectors are unit vectors, we have
\beq
w(r_1, r_2) + w(r_2, r_3) + w(r_3, r_4) + w(r_4, r_1)
= \tau^2\, (\partial_\rho W_\sigma - \partial_\sigma W_\rho)\;.
\eeq 
where $\rho$ and $\sigma$ are directions corresponding to $U$ and $V$. Remembering that $U$ and $V$ are of unit length, this can be re-written as 

\bea w(r_1, r_2) + w(r_2, r_3) + w(r_3, r_4) + w(r_4 , r_1) = \tau^2\, U^\mu\, V^\nu\, (\partial_\mu W_\nu - \partial_\nu W_\mu). \eea

Now since $U^\mu\, W_\mu = 0$ we know that
\beq
U^\mu\, V^\nu\, \partial_\nu W_\mu = - W^\mu\, V^\nu \partial_\nu U_\mu\;.
\eeq
Thus the above expression becomes 

\bea w(r_1, r_2) + w(r_2, r_3) + w(r_3, r_4) + w(r_4 , r_1) = \tau^2(p,q)\,
(U^\mu\, V^\nu\, \partial_\mu W_\nu + W^\mu\, V^\nu\, \partial_\nu U_\mu). \eea

In order to remember what we were minimizing or maximizing, this can be written in a more complete form as follows (here we used $V(\alpha(p,q)) = k_d\, \tau^d(p,q)$):

\bea & & (U^\mu\, V^\nu\, \partial_\mu W_\nu + W^\mu\, V^\nu\, \partial_\nu U_\mu) \nonumber \\
& & = (\frac{k_d}{V_0})^{1/d} \min_{V(\alpha(p,q))\geq V_0} \max_{r_1, r_2, r_3, r_4 \in \alpha(p,q)} (w(r_1, r_2) + w(r_2, r_3) + w(r_3, r_4) + w(r_4 , r_1) ) \eea 

We can now permute this equation to get another two equations:

\bea & & (V^\mu\, W^\nu\, \partial_\mu U_\nu + U^\mu\, W^\nu\, \partial_\nu V_\mu) \nonumber \\
& & = (\frac{k_d}{W_0})^{1/d} \min_{W(\alpha(p,q))\geq W_0} \max_{r_1, r_2, r_3, r_4 \in \alpha(p,q)} (u(r_1, r_2) + u(r_2, r_3) + u(r_3, r_4) + u(r_4 , r_1) ) \eea

and

\bea & & (W^\mu\, U^\nu\, \partial_\mu V_\nu + V^\mu\, U^\nu\, \partial_\nu W_\mu) \\ \nonumber 
& & = (\frac{k_d}{U_0})^{1/d} \min_{U(\alpha(p,q))\geq U_0} \max_{r_1, r_2, r_3, r_4 \in \alpha(p,q)} (v(r_1, r_2) + v(r_2, r_3) + v(r_3, r_4) + v(r_4, r_1)).  \eea

By subtracting the third equation from the sum of first two equations, and then dividing the whole thing by 2, we get

\bea & & U^\mu\, V^\nu\, \partial_\mu W_\nu = \half\, ((\frac{k_d}{V_0})^{1/d}
\min_{V(\alpha(p,q))\geq V_0} \max_{r_1, r_2, r_3, r_4 \in \alpha(p,q)} (w(r_1, r_2) + w(r_2, r_3) + w(r_3, r_4) + w(r_4 , r_1) ) \nonumber \\
& &  + (\frac{k_d}{W_0})^{1/d}
\min_{W(\alpha(p,q))\geq W_0} \max_{r_1, r_2, r_3, r_4 \in \alpha(p,q)} (u(r_1, r_2) + u(r_2, r_3) + u(r_3, r_4) + u(r_4 , r_1) ) \nonumber \\
& & -(\frac{k_d}{U_0})^{1/d}
\min_{U(\alpha(p,q))\geq U_0} \max_{r_1, r_2, r_3, r_4 \in \alpha(p,q)} (v(r_1, r_2) + v(r_2, r_3) + v(r_3, r_4) + v(r_4 , r_1) )), \eea

which is our desired term.

%\newpage
\bigskip
\noindent{\bf 2.3 Integration approach for fermions}

Lets start with $U^{\mu}V_{\mu}$ terms. 

Suppose we have two fields $U^{\mu}$ and $V^{\mu}$ given by two holonomies $u(r,s)$ and $v(r,s)$ We select an Alexandrov set given determined by poins $p \prec q$ in which the holonomies are assumed to be linear. As with other fields, we first pretend that we DO have coordinates, which we treat as uknowns, to get coordiniate-free estimate for Lagrangian density. We will then formally use that Lagrangian density as a definition of Lagrangian density of all Alexandrov sets, including non-manifoldlike. 
 
We will set time axis to go from $p$ to $q$.
\beq U_0 = \frac{u(p,q)}{\tau(p,q)}\;,\qquad
V_0 = \frac{v(p,q)}{\tau(p,q)}\;.
\eeq
For simplicity, we will just indicate $\tau(p,q) = \tau$.
\beq
\int_{\alpha(p,q)} \dd^dr\, u(p,r)\, v(p,r)
= \int_{\alpha(p,q)} (r^\mu\, r^\nu\, U_\mu\, V_\nu
- \frac{\tau^2}{2}\, U_0\, V_0) \dd^dr
= \tau^{d+2}\, (U_0\, V_0\, (I_{d0} - \frac{k_d}{4}) + U_k\, V_k\, I_{d1})\;.
\eeq
This gives us 
\bea
& &U_k\, V_k
= \frac{1}{I_{d1}} (\frac{1}{\tau^{d+2}} \int_{\alpha(p,q)} \dd^dr\, u(p,r)\,v(p,r)
- U_0\, V_0\, (I_{d0} - \frac{k_d}{4})) \\
& &=\ \frac{1}{I_{d1}}\, (\frac{1}{\tau^{d+2}} \int_{\alpha(p,q)} \dd^dr\, u(p,r)\, v(p,r) - \frac{1}{\tau^2}\, u(p,q)\, v(p,q)\, (I_{d0} - \frac{k_d}{4}))\;. \nonumber
\eea
Thus, we get
\bea
& &U^\mu\, V_\mu = \frac{1}{\tau^2}\, u(p,q)\, v(p,q) - U_k\, V_k \\
& &= \frac{1}{I_{d1}}\, \bigg(\frac{1}{\tau^2}\, (I_{d1} + \frac{k_d}{4} - I_{d0})\,
u(p,q)\, v(p,q) - \frac{1}{\tau^{d+2}} \int_{\alpha(p,q)} \dd^dr\, u(p,r)\, v(p,r)
\bigg)\;. \nonumber
\eea

Furthermore, we can use the expression for $U^{\mu}V_{\mu}$ to get the terms involving derivatives of scalar fields in a Lagrangian by replacing $v(r,s)$ with $\phi(r) - \phi(s)$:
\bea
& &U^\mu\, \partial_\mu\phi \\
& &=\ \frac{1}{I_{d1}}\, \bigg(\frac{1}{\tau^2}\, (I_{d1}
+ \frac{k_d}{4}-I_{d0})\, u(p,q)\, (\phi(q) - \phi(p)) - \frac{1}{\tau^{d+2}} \int_{\alpha(p,q)} \dd^dr\, u(p,r)\, (\phi(r)-\phi(p)) \bigg)\;. \nonumber
\eea
Now let us move to the more difficult issue of computting $e_a^{\mu} e_b^{\nu} \partial_\nu e_{c\mu}$ terms. 

Our plan is the following: 

PART 1: By using orthonormality of $e_k$, show that $e_a{}^\mu\, e_b{}^\nu\, \partial_{\nu} e_{c\mu}$ can be expressed as linear combination of the terms of the form $E_{lmn} = e_l{}^\mu\, e_m{}^\nu\, (\partial_\mu e_{n\nu} - \partial_\nu e_{n \mu} )$. This would simplify the situation tremendously since the latter somewhat resembles gauge theory which we already know how to do.

PART 2: Find a way of computting $E_{abc}$ in a coordinate-free setting of causal set. Even though, as remarked above, the resemblence to guage theory should make it easy, there are still difference with gauge theory, including the fact that we have 3 holonomies rather than 1, which makes it somewhat difficult. But orthonormality of these three holonomies will definitely help us pass through this. 

\noindent PART 1

\noindent If we expand the above expression and switch the dummy indices $\mu$ and $\nu$ on the second term we get 
\beq
E_{abc} = e_a{}^\mu\, e_b{}^\nu\, \partial_\mu e_{c\nu}
- e_a{}^\nu\, e_b{}^\mu\, \partial_\mu e_{c\nu}\;.
\eeq
If we now apply the fact that $\partial_\mu (e_a{}^\mu\, e_{c\mu}) = \partial_\mu \eta_{ab} = 0$ to the second term of above equation, we get
\beq
E_{abc} = e_b{}^\nu\, e_a{}^\mu\, \partial_\mu e_{c\nu}
+ e_c{}^\nu\, e_b{}^\mu \partial_\mu e_{a \nu}\;.
\eeq
By permuting the indices, we get 
\bea
& &E_{bca} = e_c{}^\nu\, e_b{}^\mu\, \partial_{\mu} e_{a\nu}
+ e_a{}^\nu\, e_c{}^\mu \partial_\mu e_{b \nu}
\\
& &E_{cab} = e_a{}^\nu\, e_c{}^\mu\, \partial_{\mu} e_{b\nu}
+ e_b{}^\nu\, e_a{}^\mu \partial_\mu e_{c\nu}\;.
\eea
From these expressions it is easy to see that
\beq
E_{abc} + E_{bca} - E_{cab} = 2\, e_c{}^\nu\, e_b{}^\mu\, \partial_\mu e_a{}^\nu\;.
\eeq
By switching $a$ and $c$ and dividing the expression by 2 we get
\beq
e_a{}^\nu\, e_b{}^\mu\, \partial_\mu e_c^\nu
= \frac{1}{2}\, (E_{cba} + E_{bac} - E_{acb})\;.
\eeq
Thus, the problem of computing $e_a{}^\nu\, e_b{}^\mu\, \partial_\mu e_c{}^\nu$ reduces to the problem of computing $E_{cba}$ for causal set, as desired.

\noindent PART 2

\noindent Let us first make a heuristic argument to make the best guess as to what kind of integral for us to try. 

We know from the case of electrodynamics that
\beq
e_c(p,r) + e_c (r,s) + e_c (s,p) = (r^{\mu} - p^{\mu})(s^{\nu} - p^{\nu}) (\partial_{\mu} e_{c \nu} - \partial_{\nu} e_{c \mu})\;.
\eeq
Furthermore, if we multiply the above expression by $e_a(p,r)\, e_b(p,s)$, we would get a projection onto $\mu = a$ and $\nu = b$ thus singling out $a^\mu\, b^\nu\, (\partial_\mu e_{c\nu} - \partial_\nu e_{c\mu})$ The reason we can do this easily is because we already know the coordinates of $p$ which leaves us with only one unknown coordinate. Of course, the above is just a heuristic argument so now we have to perform in order to see that it works. So we would like to compute
\beq
\int_{\alpha(p,q)} \dd^dr\, \dd^ds\, (e_c(p,r) + e_c(r,s) + e_c(s,p))\,
e_a(p,r)\, e_b(p,s)\;.
\eeq
Let us denote $\partial_{\mu} e_{c \nu} - \partial_{\nu} e_{c \mu}$ by $H_{ \mu \nu c}$. Thus, according to the current notation, 
\beq
E_{abc} = a^{\mu}b^{\nu} H_{ \mu \nu c}\;.
\eeq
Then, after we substitute
\beq
e_c(p,r) + e_c(r,s) + e_c(s,p) = (r^\mu - p^\mu)\, (r^\nu - p^\nu)\,H_{\mu\nu c}\;,
\eeq
our expression becomes
\beq
\int_{\alpha (p,q)} \dd^dr\, \dd^ds\, H_{\mu\nu c}\,
(r^\mu - p^\mu)\, (s^\nu - p^\nu)\, (r^a-p)(s^b-p^b)\;.
\eeq
If we now expand the parenthesis while dropping all the odd terms, we will get
\beq
\int_{\alpha(p,q)} \dd^dr\, \dd^ds\, (H_{\mu\nu c}\, r^\mu\, s^\nu\, r^a\, s^b
+ H_{c\mu\nu}\, r^\mu\, p^\nu\, r^a\, p^b + H_{c\mu\nu}\, p^\mu\, s^\nu\, p^a\, s^b
+ H_{\mu\nu c}\, p^\mu\, p^\nu\, p^a\, p^b)\;.
\eeq
In the first term, in order for the integration over $r$ to be non-zero, we should have $\mu =a$. Alson, in order for the integration over $s$ to be non-zero, we should have $\nu =b$. Since $a$ and $b$ are known while $\mu$ and $\nu$ are dummy indices, the first term becomes $H_{cab}\, (r^a)^2\, (s^b)^2$. Now, by substituting $p^\nu = -\frac{\tau}{2}\, \delta^\nu{}_0$ and $p^b = -\frac{\tau}{2}\, \delta^b{}_0$, the second term becomes $\frac{\tau^2}{4}\, H_{\mu 0c}\, \delta^b{}_0\, r^\mu\, r^a$. Again, in order for the integration over $r$ to be non-zero we have to set $\mu = a$ which means that the second term becomes $\frac{\tau^2}{4}\, H_{a0c}\, \delta^b{}_0\, (r^a)^2$. By the similar argument, the third term will become $\frac{\tau^2}{4}\, H_{0bc}\, \delta^a{}_0 (r^b)^2$. Finally, $p^\mu$ and $p^\nu$ in the last term will set $\mu = \nu = 0$ which would mean that $H_{\mu\nu c} = H_{00c} = 0$ by antisymmetry. So the last term drops out. Thus, the final expression becomes
\beq
\int_{\alpha (p,q)} \dd^dr\, \dd^ds\, \Big(E_{abc}\, (r^a)^2\, (s^b)^2\,
+ \frac{\tau^2}{4}\, \delta^b{}_0\, E_{a0c}\, (r^a)^2
+ \frac{\tau^2}{4} \delta^a_0 E_{0bc} (s^b)^2 \Big)\;.
\eeq
Now from dimensional analysis we know the following: 
\bea
& &\int_{\alpha(p,q)} \dd^dr = k_d\, \tau^d
\\
& &\int_{\alpha(p,q)} \dd^dr\, (r^\mu)^2 = I_{d\mu}\, \tau^{d+2}\;,
\eea
where by cylindrical symmetry $I_a = I_1$ whenever $a \neq 0$.

Substituting these we get the final answer for our integral: 
\bea
& &\int_{\alpha(p,q)} \dd^dr\, \dd^ds\, (e_c(p,r) + e_c(r,s) + e_c(s,p))\,
e_a(p,r)\, e_b(p,s) \\
& &=\ \tau^{2d+4}\, \Big(E_{abc}\, I_a\, I_b
+ \frac{k_d}{4}\, E_{a0c}\, I_a\, \delta^0{}_b
+ \frac{k_d}{4}\, E_{0bc}\, I_b\, \delta^0{}_a \Big)\;. \nonumber
\eea
Now, in the situation where both $a$ and $b$ are non-zero (hence denoted by $i$ and $j$) and remembering that by cylindrical symmetry (thus saying that $I_k = I_1$ for all $k\geq 1$ I get
\beq
\int_{\alpha(p,q)} \dd^dr\, \dd^ds\, (e_c(p,r) + e_c(r,s) + e_c(s,p))\,
e_i(p,r)\, e_j(p,s) = \tau^{2d+4}\, E_{ijc}\, I_1{}^2\;.
\eeq
This gives me 
\beq
E_{ijc} = \frac{1}{I_1^2 \tau^{2d+4}} \int_{\alpha(p,q)} \dd^dr\, \dd^ds\,
(e_c(p,r) + e_c(r,s) + e_c(s,p))\, e_i(p,r)\, e_j(p,s)\;.
\eeq
On the other hand, if we consider $a = 0$ and $b> 0$ (and to stress that $b>0$ we will denote $b$ by $k$) we get
\beq
\int_{\alpha(p,q)} \dd^dr\, \dd^ds\, (e_c(p,r) + e_c(r,s) + e_c(s,p))\,
e_0(p,r)\, e_k(p,s) = \tau^{2d+4}\, E_{0kc}\, I_1\, (I_0 + \frac{k_d}{4})\;.
\eeq
Thus, we will get
\beq
E_{0kc} = \frac{1}{\tau^{2d+4} I_1\, (I_0 + \frac{k_d}{4})} \int_{\alpha(p,q)}
\dd^dr\, \dd^ds\, (e_c(p,r) + e_c(r,s) + e_c(s,p))\, e_0(p,r)\, e_k(p,s)\;.
\eeq

Thus, the combination of results from parts 1 and 2 tells us that
\beq
e_a{}^\nu\, e_b{}^\mu\, \partial_\mu e_c^\nu
= \frac{1}{2}\, (E_{cba} + E_{bac} - E_{acb})\;,
\eeq
where
\beq
E_{ijc} = \frac{1}{I_1^2 \tau^{2d+4}} \int_{\alpha(p,q)} \dd^dr\, \dd^ds\,
(e_c(p,r) + e_c(r,s) + e_c(s,p))\, e_i(p,r)\, e_j(p,s)\;,
\eeq
when both $i$ and $j$ are non-zero, and 
\beq
E_{0kc} = -E_{k0c} = \frac{1}{\tau^{2d+4}\, I_1\, (I_0 + \frac{k_d}{4})}
\int_{\alpha(p,q)} \dd^dr\, \dd^ds\, (e_c(p,r) + e_c(r,s) + e_c(s,p))\,
e_0(p,r)\, e_k(p,s)\;.
\eeq

This determines $e_a{}^\nu\, e_b{}^\mu\, \partial_\mu e_c^\nu$.

 %\newpage
\bigskip
\noindent{\bf 2.4 Definition of f, g, and h from both approaches}
 $$
\noindent 
$$

Before we proceed to the next section, it is best to summarize the results of the previous two subsections. Since in the causal set there are no coordinates, it is not correct to say that we have found an approximation to the contractions of derivatives, since the latter is not defined. Rather, we have found functions $f(u,v;p)$, $g(u,v;p)$ and $h(u,v;p)$ that happened to have a feature that in a special case where the causal set happens to be manifoldlike, the following approximations happen to hold:  

\bea (A^{\mu}B_{\mu} )(p) \approx f_i(a,b;p) \approx f_m (a, b; p) \eea
\bea A^{\mu} \partial_{\mu} \phi \approx g_i (a, \phi; p) \approx g_m (a, \phi; p) \eea
\bea(A^{\mu} B^{\nu} \partial_{\nu} C_{\mu})(p) \approx h_i (a,b,c; p) \approx h_m (a,b,c;p) \eea

We will now rewrite down the definitions of above $f$, $g$ and $h$ which would be the key results of the previous sections, and will be used as a part of definition of Lagrangian.The $f$, $g$ and $h$ that are based on maximization/minimization approach will be denoted by $f_m$, $g_m$ and $h_m$. The $f$, $g$ and $h$ that are based on integration approach will be denoted by $f_i$, $g_i$ and $h_i$. Here, p will be a point of reference, while q(u, v; p) will be a point that would minimize fluctuations of u and v on $\alpha (p,q)$ 

\bea & & f_m (u, v; p) = \half\, \bigg(\frac{k_d}{V} \bigg)^{1/d} \Big(\min_{V(\alpha(p,q(u, v; p))) \geq V_0}\, \max_{r,s \in \alpha(p,q(u, v; p))} |u(r,s)+v(r,s)| \\
& &\kern106pt-\ \min_{V(\alpha(p,q(u, v; p))) \geq V_0}\, \max_{r,s \in \alpha(p,q(u, v; p))} |b(r,s)| - \min_{V(\alpha(p,q(u, v; p))) \geq V_0}\, \max_{r,s \in \alpha(p,q(u, v; p))} |c(r,s)|\Big)\,.\nonumber
\eea

\bea
& & g_m (v, \phi) = \half\, \bigg(\frac{k_d}{V} \bigg)^{\!1/d} \Big(\min_{V(\alpha(p,q))\geq V_0}\,\max_{r,s\in\alpha(p,q)}|v(r,s)+\phi(s)-\phi(r)|\\
& &\kern108pt-\ \min_{V(\alpha(p,q)) \geq V_0}\, \max_{r,s \in \alpha(p,q)} |v(r,s)| - \min_{V(\alpha(p,q)) \geq V_0}\, \max_{r,s \in \alpha(p,q)} |\phi(s)-\phi(r)|\Big)\,.
\nonumber
\eea

\bea & & h_m (u, v, w) = \half\, ((\frac{k_d}{V_0})^{1/d}
\min_{V(\alpha(p,q))\geq V_0} \max_{r_1, r_2, r_3, r_4 \in \alpha(p,q)} (w(r_1, r_2) + w(r_2, r_3) + w(r_3, r_4) + w(r_4 , r_1) ) \nonumber \\
& &  + (\frac{k_d}{W_0})^{1/d}
\min_{W(\alpha(p,q))\geq W_0} \max_{r_1, r_2, r_3, r_4 \in \alpha(p,q)} (u(r_1, r_2) + u(r_2, r_3) + u(r_3, r_4) + u(r_4 , r_1) ) \nonumber \\
& & -(\frac{k_d}{U_0})^{1/d}
\min_{U(\alpha(p,q))\geq U_0} \max_{r_1, r_2, r_3, r_4 \in \alpha(p,q)} (v(r_1, r_2) + v(r_2, r_3) + v(r_3, r_4) + v(r_4 , r_1) )), \eea

\bea
& &f_i (u,v; p)   = \frac{1}{I_{d1}}\, \bigg(\frac{1}{\tau^2}\, (I_{d1} + \frac{k_d}{4} - I_{d0})\,
u(p,q(u,v,p))\, v(p,q(u,v;p)) \nonumber \\
& & - \frac{1}{\tau^{d+2}} \int_{\alpha(p,q(u,v;p))} \dd^dr\, u(p,r)\, v(p,r)
\bigg)\; 
\eea

\bea
& &g_i (v, \phi ; p)  =\ \frac{1}{I_{d1}}\, \bigg(\frac{1}{\tau^2}\, (I_{d1}
+ \frac{k_d}{4}-I_{d0})\, u(p,q(\phi; p))\, (\phi(q(\phi;p)) - \phi(p))  \nonumber \\
& & - \frac{1}{\tau^{d+2}} \int_{\alpha(p,q(\phi;p))} \dd^dr\, u(p,r)\, (\phi(r)-\phi(p)) \bigg)\;
\eea

 \beq
h_i (a, b, c; p ) = \frac{1}{2}\, (E(c,b,a;p)  + E(b,a,c;p)- E(a,c,b;p) )\;,
\eeq

where

\beq
E(b,c,d;p )= \frac{1}{I_1^2 \tau^{2d+4}} \int_{\alpha(p,q(b,c,d;p))} \dd^dr\, \dd^ds\,
(d(p,r) + d(r,s) + d(s,p))\, b(p,r) \, c(p,s)\;,
\eeq
when $f(b,b)<0$, $f(c,c)<0$ and $f(d,d)<0$ (which is a causal set way of saying that all three holonomies correspond to spacelike vectors)

and 

\beq
E(a,b,c) = -E(b,a,c) = \frac{1}{\tau^{2d+4}\, I_1\, (I_0 + \frac{k_d}{4})}
\int_{\alpha(p,q)} \dd^dr\, \dd^ds\, (c(p,r) + c(r,s) + c(s,p))\,
a(p,r)\, b(p,s)\;.
\eeq

if $f(a,a)>0$, while $f(b,b)<0$ and $f(c,c)<0$ (which is a causal set way of saying that a corresponds to a timelike vector while b and c correspond to spacelike vectors).

\bigskip
\noindent{\bf 3. Causal set verion of Ref \cite{paper4} }

In the previous section, we have found a way to write $A^{\mu}B_{\mu}$, $A^{\mu} \partial_{\mu} \phi$ and $A^{\mu} B^{\nu} \partial_{\nu} C_{\mu}$ as functions of holonomies corresponding to these vector fields,  

\bea (A^{\mu}B_{\mu} )(p) \approx f_i(a,b;p) \approx f_m (a, b; p) \eea
\bea A^{\mu} \partial_{\mu} \phi \approx g_i (a, \phi; p) \approx g_m (a, \phi; p) \eea
\bea(A^{\mu} B^{\nu} \partial_{\nu} C_{\mu})(p) \approx h_i (a,b,c; p) \approx h_m (a,b,c;p) \eea

where indeces $i$ and $m$ correspond to maximization/minimization approach while index $i$ corresponds to integration approach. From now on we will drop these indeces and just use $f$, $g$ and $h$. Reader can insert either of these indeces depending on their preference of which approach to use. 

We will now use $f$, $g$ and $h$ to write down a complete fermionic Lagrangian. 

In order to make notation more intuitive, we will define dot products of holonomies (which is not to be confused with a dot product used on Grassmann space) as follows:

\bea (a \cdot b) (p) = f(a,b; p)) \eea
\bea ((a \cdot b) c) (p,q) = (a \cdot b) (p) c(p,q) \eea

Thus, in a special case of a manifold, 

\bea (a \cdot b)(p) = (A^{\mu} B_{\mu})(p) = g^{\mu \nu} \frac{\partial a(p,r)}{\partial r^{\mu}}\frac{\partial b(p,s)}{\partial s^{\nu}} \eea

and 

\bea ((a \cdot b) c)(p,q) =  (A^{\mu} B_{\mu})(p) C_{\nu} (q^{\nu} - p^{\nu}) = g^{\mu \nu} \frac{\partial a(p,r)}{\partial r^{\mu}}\frac{\partial b(p,s)}{\partial s^{\nu}} c(p,q) \eea

Now, in the same way as $A^{\mu}$ through $D^{\mu}$ were just described as holonomies, we will also introduce a modified Gramm Schmidt process that would define orthogonal basis $e_a^{\mu}$ in terms of holonomies as well:

\bea (e_0 (a,b)) (r,s)= \frac{a(r,s)}{\sqrt{(a \cdot a) (r)}} \eea
\bea (e_1(a,b)) (r,s)= \frac{b(r,s) - e_0 (r,s) (e_0 \cdot b))(r)}{\sqrt{((b-(e_0 \cdot b) e_0) \cdot (b-(e_0 \cdot b) e_0)) (r) }} \eea
\bea (e_2 (a,b)) (r,s)= \frac{c(r,s) - e_0 (r,s) (e_0 \cdot c) - e_1 (r,s) (e_1 \cdot c))(r)}{\sqrt{((c-(e_0 \cdot c) e_0- (e_1 \cdot c) e_1) \cdot (c-(e_0 \cdot c) e_0 - (e_1 \cdot c) e_1)) (r) }} \eea
\bea (e_3 (a,b)) (r,s)= \frac{d(r,s) - e_0 (r,s) (e_0 \cdot d) - e_1 (r,s) (e_1 \cdot d)- e_2 (r,s) (e_2 \cdot d))(r)}{\sqrt{((d-(e_0 \cdot d) e_0- (e_1 \cdot d) e_1- (e_2 \cdot d) e_2) \cdot (d-(e_0 \cdot d) e_0 - (e_1 \cdot d) e_1- (e_2 \cdot d) e_2)) (r) }} \eea

We will also define a basis $f_a^{\mu}$ such that $f_0$ through $f_3$ are unit vectors in the directions $A$ through $D$:

\bea & & f_0 (a) (p,q) = \frac{a(p,q)}{\sqrt{(a \cdot a)(p)}} , \;  f_1(b) (p,q) = \frac{b(p,q)}{\sqrt{(b \cdot b)(p)}} , \nonumber \\
& & f_2 (c) (p,q) = \frac{c(p,q)}{\sqrt{(c \cdot c)(p)}} , \; f_3 (d) (p,q) = \frac{d(p,q)}{\sqrt{(d \cdot d)(p)}} \eea

From this point on we can pursue two different models, which are different from each other in terms of the extend to which we apply the context of Ref \cite{paper4}. In both models we do have to use what we refer in the current paper as "part 1" of Ref \cite{paper4} that deals with literal definition of Grassmann numbers since we need that in order to analyze the fluctuations of the latter. As was mentioned earlier, even if we choose the integration approach, we would still have to analyze the fluctuations of fields in the interior of Alexandroff sets in order not to select an Alexandroff set that lies near the lightcone which might result in singular behavior due to nonlinearity of fields. However, while we do have to use what we refer in current paper as "part 1" of Ref \cite{paper4} , we don't have to use parts 2 and 3 of Ref \cite{paper4} . We might instead do the same thing as is done in standard quantum field theory: view $\psi_i$ as degrees of freedom independent of vierbines or any other vector fields, and simply couple them to vierbines. Thus, we have two options:

OPTION 1: Use only what current paper calls "part 1" of Ref \cite{paper4} 
OPTION 2: Use all three parts of Ref \cite{paper4}

Both points of view have their own advantages and disadvantages.

Advantages of options 1:

a)  The measure $\lambda$ involved in option 2 is a lot more complicated than the measure $\xi$ that will be used in option 1 because of the need to take "derivative" of fermionic degrees of freedom with respect to vectors that define them.

b)  Since in order to talk about fermions we are forced to introduce vierbines in one way or the other, it is tempting to use them for a different purpose as well: to adress the question as to why causal set manifoldlike. As will be explained in more details in the next section, in order for holonomies related to vierbines to be of use in that respect, we would need to use random distribution, i.e. measure has to be constant. This is not possible to do in option 2 since the Grassmann measure $\lambda$ used in the latter takes into account both $\psi$ as well as holonomies, and $\lambda$ is definitely uneven.  In option 1, on the other hand, we can still do that since there holonomies and vector fields are treated as separate. 

Advantages of option 2: 

a) If vierbines  are treated as separate from the spinor field, they feel like they were put in by hand. This make it seem like a certain direction is "better" than others since spin up and spin down are taken with respect to that one directoin. The "better" direction is, of course, given by vierbines. 

From philosophical point of view, option 2 is clearly more advisable. However, from computational point of view option 1 is better since, as you will find, the measure $\lambda$ involved in option 2 is a lot more complicated than the measure $\xi$ that will be used in option 1 because of the need to take "derivative" of fermionic degrees of freedom with respect to vectors that define them. Since the ultimate way of testing causal set theories is numerical methods, the computational aspect should not be ignored. However, since part of my motivation of doing causal sets on the first place is philosophy, the philosophical aspect should not be ignored either. For this reason, I resort of doing two different models and separately doing option 1 and option 2. 

 %\newpage
\bigskip
\noindent{\bf  Option 1}
 $$
\noindent 
$$

 In Option 1, Dirac field $\psi$ will be interpretted as 4 independent complex scalar fields.  Since holonomies $a$, $b$, $c$ and $d$ determines vierbines, I will say that these holonomies will be interacting with the four components of spinor field I have just introduced. 

The derivative of above defined spinor field is given by

\bea \partial_m \psi_i = g(e_m , \psi_i , \alpha (p,q)) \eea

The gauge field with which it interacts (such as for example electromagnetic field) is given by a holonomy $v (r,s)$ which means that in a frame given by our vierbines, 

\bea V^m = f(e^m, v, \alpha(p,q)) \eea

The interaction of fermionic and gauge field is given by 

\bea \overline{\psi} \gamma^{\mu} V_{\mu}  \psi = \psi_i^* \psi_j (\gamma^0 \gamma^{\mu})_{ij} f(e^m, v, \alpha (p,q)) \eea

This means that the Lagrangian is given by

\bea & &  {\cal L} (a, b, c, d)(p)= (\gamma^0 \gamma^{\mu})_{ij} \psi_i^* (p) g(e_m (a, b, c, d), \psi_j ) (p) 
\nonumber \\
& & + \psi_i ^*(\chi_p , \chi_a, a, b, c, d) (p)  \psi_j (p) (\gamma^0 \gamma^m \sigma^{ab})_{ij} (h(e_a (a, b, c, d), e_b (a, b, c, d), e_b(a, b, c, d)) (p) 
\nonumber \\
& & - h(e_m (a, b, c, d), e_a (a, b, c, d), e_b (a, b, c, d) )(p)  + h(e_a (a, b, c, d), e_b(a, b, c, d), e_m (a, b, c, d)) (p)) \nonumber \\
& & + \psi_i^* (p) \psi_j  (\gamma^0 \gamma^{\mu})_{ij} f(e^m (a, b, c, d), v) (p) \eea

Finally, the amplitude is given by

\bea & & Z = \int [da] [db] [dc] [dd] [d \psi_1] [d \psi_2] [d \psi_3] [d \psi_4] \; \xi (Re (\psi_1) ) \xi (Im (\psi_1)) \xi (Re (\psi_2)) \xi (Im (\psi_2)) \times \nonumber \\
& & \times \xi(Re (\psi_3) ) \xi (Im (\psi_3)) \xi (Re (\psi_4) ) \xi (Im (\psi_4) )  (\hat{r_1} \wedge . . . \wedge \hat{r_8}) \cdot e^{iS (\psi))}  \eea

 %\newpage
\bigskip
\noindent{\bf  Option 2}
 $$
\noindent 
$$

In this option, $\psi$ will be a function of vector fields, which in causal set case means holonomies. We simply rewrite the definition of $\psi$ in terms of vector fields given in \cite{paper4} and repeated in equations (14) and (15) and rewrite it in the language of causal sets. This would give us the following:

\bea
& & \psi_i(\chi_p , \chi_a , a, b, c, d) (p)\\
& & = \big(\exp\big\{ -{\textstyle\frac{\ii}{4}} (\ln (e^{-1} (a, b, c, d)(p) f (a, b, c, d)(p)))_{\mu \nu} \sigma^{\mu\nu} \big\}\, \big)_{ij} \big(\chi_p (p) \delta^j_1  + \chi_a (p) \delta^j_3  \big) \;. 
\nonumber
\eea

Likewise, the measure on $\psi$ space is nothing but a copy of equation (16) in causal set language: 

\bea & & \lambda (\chi_p , \chi_a , a, b, c, d) =\xi [Re (\psi (\chi_p, \chi_a, a, b, c, d ))] \xi [Im (\psi (\chi_p, \chi_a, a, b, c, d ))]\times \\ \nonumber
& &  \times lim_{\epsilon \rightarrow 0} \epsilon \mu^{-1} \Big\{ \chi_p ' , \chi_a ', a' , b' , c' , d' \Big\vert \big\vert \psi (\chi_p ', \chi_a ', a', b', c', d'  ) \nonumber \\
& & - \psi (\chi_p , \chi_a , a, b, c, d ) \vert < \epsilon  \wedge \forall a [(e_a^{\mu} ( a', b', c', d' )) - e_a^{\mu} ( a, b, c, d ) )  \nonumber \\
& & (e_{a\mu} (a',b',c',d')   - e_{a\mu} (a,b,c,d)) ] < \epsilon^2 \Big\}  \nonumber \eea 

where $\mu$ is a usual Eucledian measure on $\mathbb{R}^{18}$ 

The Lagrangian density is given by rewriting equation (93) while inserting the definition of $\psi$ : 

\bea & &  {\cal L} (a, b, c, d)(p)= (\gamma^0 \gamma^{\mu})_{ij} \psi_i^* (\chi_p, \chi_a, a, b, c, d)(p) g(e_m (a, b, c, d), \psi_j (\chi_p, \chi_a, a, b, c, d)) (p) 
\nonumber \\
& & + \psi_i ^*(\chi_p , \chi_a, a, b, c, d) (p)  \psi_j (\chi_p, \chi_a, a, b, c, d) (p) (\gamma^0 \gamma^m \sigma^{ab})_{ij} (h(e_a (a, b, c, d), e_b (a, b, c, d), e_b(a, b, c, d)) (p) 
\nonumber \\
& & - h(e_m (a, b, c, d), e_a (a, b, c, d), e_b (a, b, c, d) )(p)  + h(e_a (a, b, c, d), e_b(a, b, c, d), e_m (a, b, c, d)) (p)) \nonumber \\
& & + \psi_i^* (\chi_p , \chi_a , a, b, c, d)(p) \psi_j (\chi_p , \chi_a , a, b, c, d) (\gamma^0 \gamma^{\mu})_{ij} f(e^m (a, b, c, d), v) (p) \eea

Finally, the amplitude is

\bea & & Z = \int [da] [db] [dc] [dd] [d \chi_p] [d \chi_a] \; \lambda (\chi_p, \chi_a, A,B,C,D)  \lambda (\chi_p, \chi_a, A,B,C,D) \nonumber \\
& & \times \lambda (\chi_p, \chi_a, A,B,C,D) \lambda (\chi_p, \chi_a, A,B,C,D)  (\hat{r_1} \wedge . . . \wedge \hat{r_8}) \cdot e^{iS (\vec{\psi} (\chi_p, \chi_a, A,B,C,D))}  \eea

where $S (\psi) = \int {\cal L} (\psi) d^d x$ is defined in the usual way, where $\wedge$ product is used for multiplication. 

\bigskip
\noindent{\bf 4. Discussion}

This paper completted the main steps of introducing matter fields into causal set theory. Scalar fileds were introduced in Ref \cite{paper1} and Ref \cite{paper5}. Gauge fields were introduced in Ref \cite{paper3} and Ref \cite{paper5} and gravitational field was introduced in Ref \cite{paper1}. The goal of this paper is to similarly introduce fermionic fields into causal set theory. 

However, introducing fermionic fields into causal set theory does more than to simply complete a list of fields. In particular, there is an important philosophical result of what we mean by fermions in general: namely that fermions can be interpreted as local frame defined by non-orthonormal vector fields. While similar thing was implied by \cite{paper4} as well, the lack of aforegiven manifold structure in this paper makes this point stronger. In fact, fermions can be viewed as agents that cause causal set to be manifoldlike, which is an interesting observation since the issue of manifoldlike-ness of causal set is the unresolved one by this point. However, one has to be a little more careful. In principle, for any partial order it is possible to adjust a set of vierbines in such a way that $e_0^2 (p,q) - e_1^2 (p,q) - e_2^2 (p,q) - e_3^2 (p,q)$ is positive if and only if $p$ and $q$ are causally related. The key thing here is the fact that while such choices do exist, if causal set is non-manifoldlike, they are "rare". Thus, since integration over holonomies is restricted to the ones that satisfy above condition, the integral is small since the range of integration is small. This would imply that manifoldlike causal sets are "selected out" as more probable ones. 

The above argument implicitly relies on the fact that measure is uniform in the space of holonomies. This would not be true for the "option 2" approach (see above) since in such an approach vierbeins, being "mixed" with spinor fields, are being subject to measure that has both positive and negative values. Furthermore, the measure is defined really based on the measure of a range of the map from holonomy space to $\psi$ space. Thus, if "a lot" of holonomies are mapped into the same values of $\psi$, the fact that there are a lot of them won't make measure any larger. This means that the fact that in case of non-manifoldlike causal set only "few" holonomies would satisfy the restrictions does not imply that the corresponding measure is small and thus does not imply the smallness of the integral. This means that if we are to use fermions in order to explain the manifoldlikeness of causal set, we are to stick to option 1.

However, option 1 gives us different questions. Since in option 1 fermions and vierbines are separate, vierbines are no longer fields. But a vierbine is precisely what tells us  what direction is the one with respect to which spin might be ÓupÓ or ÓdownÓ. So if vierbines are not Þelds, then the question is what does make that direction ÓbetterÓ? One answer is to go back to more standard view in which they are still viewed as fields, but instead of being part of fermionic field they are square roots of gravitational field. Incidentally, as already noticed by others, this would have an advantage that instead of taking square root of determinant of a metric, we would have determinant of vierbine without square root, which might possibly adress renormalization problem. 

There are two problems with this approach, however. First of all, if we take this view seriously, then we would have to abandon causal structure altogether, as opposed to emposing a constraint on consistency of vierbine related holonomies with causal structure. In this case we would be forced to adress the question of quantum gravity: how can gravitational field be both a stage and an actor? Of course, such question has to be adressed anyway, which is the ultimate point of causal set theory.  But until it is adressed, we would like to be able to be able to describe fermionic quantum field theory on a fixed gravitational background. It might still be useful from philosophical point of view to have fixed causal set background as opposed to fixed manifold background since I would like to claim that fermions are ultimately responsible for causal structure. In this case, the fact that vierbines vary while causal structure is fixed might be an argument against viewing vierbines as part of gravitational field, especially in light of the fact that option 2 offers an alternative. Although still, in option 1 this can be avoided by saying that vierbines do define gravitational field, and they simply vary within a certain constraint that $e_0^2 (p,q) - e_1^2 (p,q) - e_2^2 (p,q) - e_3^2 (p,q)$ either stays positive or stays negative for any pair of points. 

Another problem with option 1 is that if vierbines are to be viewed as fields, they would contradict any other field that we know in a sense that their variation is severely restricted by orthogonality and norm $1$. This can be adressed through Lagrange multipliers, but viewing Lagrange multipliers as fields would only pose more problems. This brings us to the Gramm Schmidt process of part 2. This allows us to define vierbines not as fields themselves but rather as functions of other fields, and behavior of the latter is not constraint. But in this case we would have to worry about the measure due to the fact that we would have to take "derivative" of vierbines with respect to original vectors. In this case, one can argue that since we have to introduce measure anyway, we might as well introduce the kind of measure that is usefull for Grassmann integral, which would bring us right back to option 2. Furthermore, since Gramm Schmidt process is not the only way of selecting orthonormal basis, one can argue that it is "arbitrary" anyway, and if it is, why can't we "arbitrarily" use the remaining degrees of freedom for the definition of fermionic field as is done in option 2? Of course, as was said earlier, the price to pay for option 2 is that due to the exotic measure we can no longer use fermions to explain the manifoldlike structure of causal set. But one can claim that lack of manifoldlike structure is a problem that has always been there, so we haven't created any new problems; we simply didn't adress as many problems as we hoped. Nevertheless, option 2 DOES define fermions as local frames; it simply doesn't adress the question of manifoldlikeness, which means, for example, smooth transition between these frame.

%\newpage

\end{document}